\documentclass[conference]{IEEEtran}
\IEEEoverridecommandlockouts
\usepackage{cite}
\usepackage{amsmath,amssymb,amsfonts}
\usepackage{algorithmic}
\usepackage{graphicx}
\usepackage{booktabs}
\usepackage{amsmath}
\usepackage{textcomp}
\usepackage{xcolor}
\usepackage{hyperref}
\usepackage{url}
\usepackage{booktabs}
\usepackage{tabularx}
\def\BibTeX{{\rm B\kern-.05em{\sc i\kern-.025em b}\kern-.08em
    T\kern-.1667em\lower.7ex\hbox{E}\kern-.125emX}}

\begin{document}

\title{
LaGDif: Latent Graph Diffusion Model for Efficient Protein Inverse Folding with Self-Ensemble
}

\author{\IEEEauthorblockN{Taoyu Wu\textsuperscript{1}, Yu Guang Wang\textsuperscript{2,3},
Yiqing Shen\textsuperscript{3,4,*}}
\IEEEauthorblockA{
\textsuperscript{1}\textit{Department of Computer Science}, \textit{University of Ottawa}, Ottawa, Canada \\
\textsuperscript{2}\textit{Institute of Natural Sciences}, \textit{Shanghai Jiao Tong University}, Shanghai, China\\
\textsuperscript{3}\textit{Toursun Synbio}, Shanghai, China \\
\textsuperscript{4}\textit{Department of Computer Science}, \textit{Johns Hopkins University}, Baltimore, MD, USA \\
\textsuperscript{*}Denotes the corresponding author. \qquad 
twu045@uottawa.ca\qquad yiqingshen1@gmail.com
}
}

\maketitle

\begin{abstract}
Protein inverse folding aims to identify viable amino acid sequences that can fold into given protein structures, enabling the design of novel proteins with desired functions for applications in drug discovery, enzyme engineering, and biomaterial development. 
Diffusion probabilistic models have emerged as a promising approach in inverse folding, offering both feasible and diverse solutions compared to traditional energy-based methods and more recent protein language models.
However, existing diffusion models for protein inverse folding operate in discrete data spaces, necessitating prior distributions for transition matrices and limiting smooth transitions and gradients inherent to continuous spaces, leading to suboptimal performance.
Drawing inspiration from the success of diffusion models in continuous domains, we introduce the Latent Graph Diffusion Model for Protein Inverse Folding (LaGDif).
LaGDif bridges discrete and continuous realms through an encoder-decoder architecture, transforming protein graph data distributions into random noise within a continuous latent space.
Our model then reconstructs protein sequences by considering spatial configurations, biochemical attributes, and environmental factors of each node. 
Additionally, we propose a novel inverse folding self-ensemble method that stabilizes prediction results and further enhances performance by aggregating multiple denoised output protein sequence.
Empirical results on the CATH dataset demonstrate that LaGDif outperforms existing state-of-the-art techniques, achieving up to 45.55\% improvement in sequence recovery rate for single-chain proteins and maintaining an average RMSD of 1.96 \AA between generated and native structures. 
These advancements of LaGDif in protein inverse folding have the potential to accelerate the development of novel proteins for therapeutic and industrial applications. 
The code is public available at \url{https://github.com/TaoyuW/LaGDif}.
\end{abstract}


\section{Introduction}
Protein inverse folding, by predicting amino acid (AA) sequences that can fold into a given protein structure, is a fundamental challenge in computational biology with far-reaching implications for drug discovery, enzyme engineering and \textit{etc} \cite{zhou2023prorefiner}.
The ability to design proteins with specific structural features is important for designing novel proteins with desired functions, such as membrane proteins for targeted drug delivery or enzymes for industrial applications \cite{janes2024deep}.
Recent advancements in deep learning have  substantially improved protein structure modeling by automatically learning complex nonlinear relationships from protein data \cite{ferruz2022protgpt2}, demonstrating success in predicting protein structures from sequences \cite{lategan2023seqprednn}.
In parallel, current approaches to protein inverse folding rely on language models that treat the 20 types of AAs as tokens, effectively transforming the task into a sequence generation problem analogous to natural language processing \cite{lin2023evolutionary,hsu2022learning,mcpartlon2022deep}.
These language-based models typically fall into two categories, namely the masked models and autoregressive models \cite{bepler2021learning}.
Masked models predict unknown AAs by leveraging contextual information from partially masked sequences, while autoregressive models sequentially predict the probability distribution of each subsequent AA based on the existing sequence \cite{anand2022protein}.
However, language models are limited to capture the inherent one-to-many mapping between protein structures and their corresponding AA sequences \cite{yi2024graph}.
The deterministic nature of these models, even when incorporating randomness through techniques like temperature sampling, struggles to fully represent the diversity of possible sequences that can fold into a given structure.

Diffusion probabilistic models have emerged as a promising alternative to traditional language-based models for protein sequence generation, offering capabilities in modeling and generating diverse outputs \cite{anand2022protein}.
These diffusion models generate data through a reverse process from noise to structured data, possessing a natural advantage in handling uncertainty and exploring complex, multimodal output spaces \cite{watson2023novo}.
This characteristic makes them therefore well-suited for accurately capturing the relationship between protein structures and multiple possible AA sequences.
Recent work has demonstrated the potential of diffusion models in protein inverse folding. 
For instance, GradIF \cite{yi2024graph} introduced a discrete diffusion model for protein inverse folding, utilizing the BLOSUM matrix as a transition matrix. 
However, discrete diffusion models operate in a discrete space, which can lead to less smooth distributions of generated samples compared to continuous space models \cite{yang2023fast}.
It can affect the quality and diversity of the generated protein sequences.
Furthermore, many sampling scheme, such as self-ensemble \cite{shen2023staindiff}, are challenging to apply to discrete diffusion models, potentially limiting the efficiency of the generation process.
To address these limitations, we propose a novel approach that constructs a diffusion model in the AA latent space, employing a graph encoder-decoder architecture to transform discrete AAs into a continuous latent representation.

The major contributions are four-fold. 
First, we introduce LaGDif, a latent space diffusion model based on Equivariant Graph Neural Network (EGNN) \cite{satorras2021n}, which excels in generating diverse protein sequences while maintaining structural integrity.
LaGDif addresses the limitations of discrete diffusion models by operating in a continuous latent space, allowing for smoother distributions of generated samples and improved exploration of the sequence space.
Secondly, to enhance the biological relevance of sample distributions in latent space, we leverage the ESM2 pretrained encoder \cite{lin2023evolutionary} for AA encoding, incorporating valuable evolutionary information into our model. 
Additionally, we integrate secondary structure information using the DSSP method, enabling a more comprehensive understanding of protein spatial configuration.
Thirdly, we propose a guided sampling approach with controlled noise, which ensures diversity in generated sequences while preserving essential structural integrity. 
This method facilitates a more efficient and stable denoising process by initializing from a point closer to the target distribution.
Finally, we introduce a novel inverse folding self-ensemble method that stabilizes prediction results and further enhances performance, which aggregates multiple denoised output protein sequences at each sampling step, leading to more robust and accurate predictions by mitigating individual biases and errors.

\begin{figure}[t!]
    \centering
    \includegraphics[width=\linewidth]{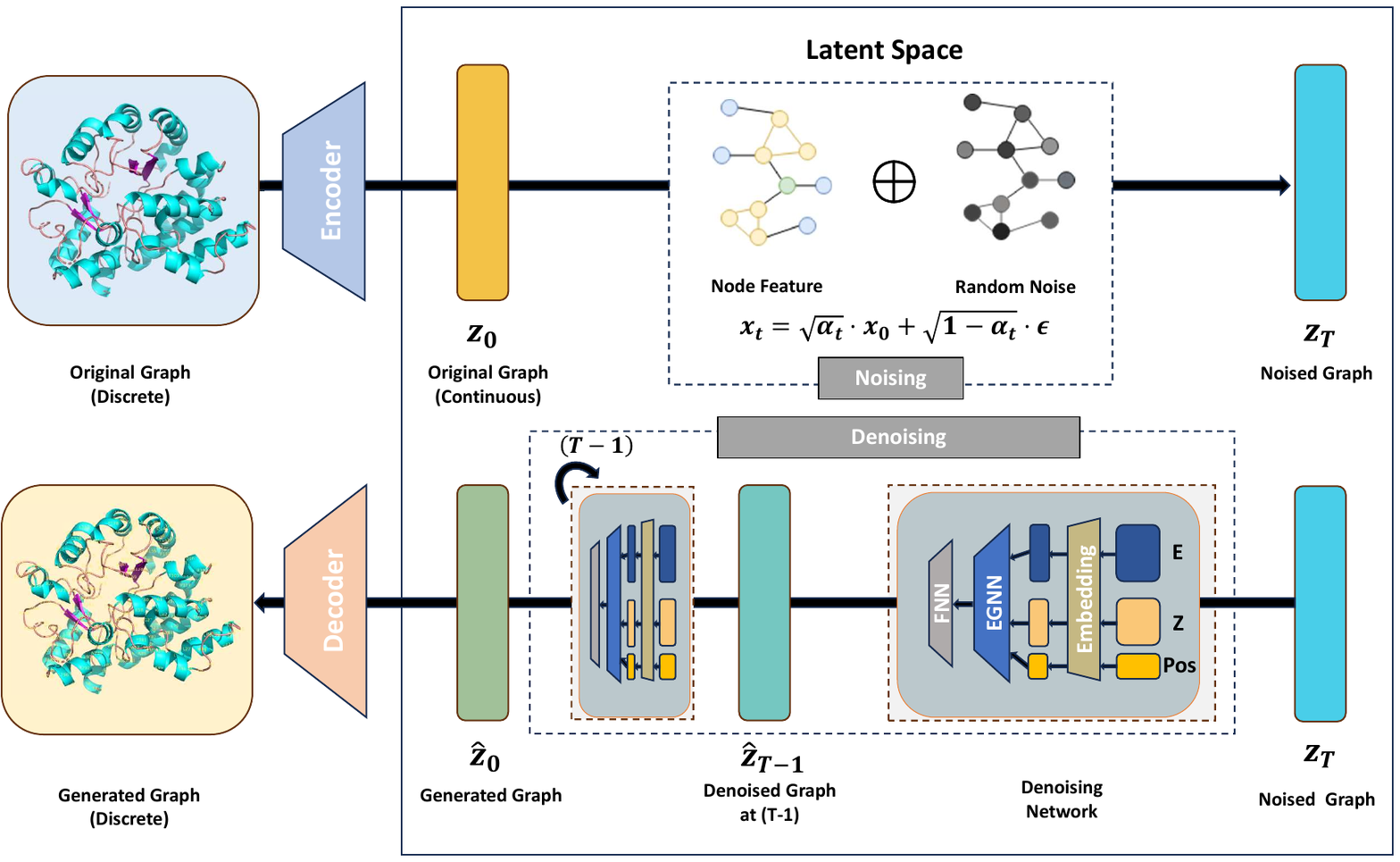}
    \caption{Overview of LaGDif. 
    The upper path illustrates the forward process, where the original protein graph is encoded into a latent space and progressively diffused with Gaussian noise. 
    The lower path shows the reverse process, where a noised graph is iteratively denoised using an equivariant neural network, conditioned on structural information, to generate diverse protein sequences compatible with the given structure. 
    The encoder-decoder architecture, based on ESM2, facilitates the transition between discrete amino acid sequences and the continuous latent space, enabling the application of diffusion models to protein inverse folding.}
    \label{fig:method}
\end{figure}

\section{Methods}
In this section, we elaborate on LaGDif, a latent graph diffusion model for protein inverse folding, as shown in Fig.~\ref{fig:method}.

\subsection{Transition to Continuous Space}
Given a protein graph $G = (V, E)$, where $V$ represents the feature matrix of the nodes and $E$ is the adjacency matrix, and $X_{pos}$ denotes the three-dimensional positional information of nodes, the encoding and decoding processes are defined as:
\begin{equation}
\begin{aligned}
    z &= \text{Encoder}(V, X_{pos}, E)\\
    \hat{x} &= \text{Decoder}(z, X_{pos}, E).\label{eq:1}
\end{aligned}
\end{equation}
Our encoder is based on the ESM2 \cite{lin2022language}, a state-of-the-art protein language model pre-trained on a vast corpus of protein sequences. 
By leveraging ESM2, our encoder benefits from a deep understanding of AA relationships and protein structure, capturing rich semantic information from the context, allowing for a more informative and biologically relevant latent representation.
The encoder transforms discrete amino acid sequences into a high-dimensional continuous latent space, preserving crucial structural and biochemical information.
The decoder, implemented as a learnable linear layer, efficiently maps the high-dimensional continuous latent representations back to discrete AA sequences. 

\subsection{Latent Graph Diffusion Model}

\paragraph{Diffusion Model.} 
The diffusion model in LaGDif is a generative probabilistic model that learns to denoise protein graphs.
It is trained on continuous protein graphs that are gradually subjected to noise, effectively capturing the distribution of AA in latent space. 
Through this process, the diffusion model learns to reverse the noise addition, allowing it to generate protein sequences by iteratively denoising random noise back into the sample space of AAs. 
This approach enables the exploration of diverse, yet structurally valid protein sequences with the randomness during the sampling stage.

\paragraph{Modeling in Latent Space.}
We employ ESM-2 \cite{lin2022language} to compress discrete node features from the protein graph into a more efficient continuous latent space. 
This abstraction serves multiple purposes. It imbues the feature vectors with richer semantic information, capturing complex relationships between AAs. 
Additionally, it eliminates the computational burden associated with high-frequency repetitive features, allowing the diffusion model to focus on critical semantic positions. 
The continuous nature of the latent space introduces inherent smoothness, enabling the application of various conditioning mechanisms. 
%

\paragraph{Equivalent Graph Neural Network (EGNN) as Denosing Network.}
The three-dimensional structure of proteins exhibits spatial equivariance, meaning that the properties of a protein molecule should remain consistent regardless of changes in its spatial position or orientation \cite{roche20233}. 
To model this characteristic, we employ an Equivalent Graph Neural Network (EGNN) as the denoising network, which leverages the rotational and transnational equivalence of neural networks \cite{satorras2021n}.
The EGNN processes the protein graph through a series of Equivalent Graph Convolution Layers (EGCL) \cite{satorras2021n}. 
Each EGCL takes as input a set of n hidden node states $H^l = {h_1^l, h_2^l, \dots, h_n^l}$, edge embedding $m_{ij}^l$ for interconnected nodes i and j, and spatial coordinates of the nodes $X_{\text{pos}}^l = {x_{1_{\text{pos}}}^l, x_{2_{\text{pos}}}^l, \dots, x_{n_{\text{pos}}}^l}$. 
Formally, the EGCL operation can be expressed as:
\begin{equation}
H^{l+1}, X_{\text{pos}}^{l+1} = \text{EGCL}(H^l, X_{\text{pos}}^l, M).
\end{equation}
Within each EGCL, the following equations define the update process:
\begin{equation}
\begin{aligned}
m_{ij} &= \phi_e \left(h_i^l, h_j^l, |x_i^l - x_j^l|^2, m_{ij}\right) \\
x_i^{l+1} &= x_i^l + \frac{1}{n} \sum_{j \neq i} (x_i^l - x_j^l) \phi_x(m_{ij}) \\
h_i^{l+1} &= \phi_h \left(h_i^l, \sum_{j \neq i} m_{ij} \right)
\end{aligned}
\end{equation}
Here, $\phi_e$, $\phi_x$, and $\phi_h$ are learnable functions implemented as neural networks, and the term \( \frac{1}{n} \) normalizes the coordinate updates based on the number of nodes to ensure stability.
The edge features $m_{ij}$ are updated based on the node features, their squared distance, and previous edge features. The node positions $x_i$ are updated using the relative positions and a learned function of the edge features. Finally, the node features $h_i$ are updated based on their previous values and aggregated edge information. 
By leveraging the EGNN as the denoising network in our latent graph diffusion model, LaGDif can effectively capture and utilize the spatial relationships between AAs, leading to more accurate and structurally sound protein sequence predictions in the inverse folding.

\subsection{Prior Knowledge of Protein Observation}
Proteins exhibit a hierarchical structural organization that are important in their function and folding process. 
The primary structure, defined by the linear sequence of amino acids, forms the foundation of protein architecture and determines the fundamental properties of the protein chain.
Building upon this, the secondary structure comprises recurring local patterns within the protein backbone, arising as a direct consequence of the primary sequence and its interactions.
To leverage the secondary structural information in our LaGDif, we employ the Definition of Secondary Structure of Proteins (DSSP) method \cite{kabsch1983dictionary}. 
DSSP analyzes the three-dimensional structure of proteins and classifies each AA into one of eight distinct secondary structure categories.  

We represent this information using one-hot encoding, and after processing it through an embedding layer, we integrate it with other features into the diffusion model's input.
%
%

\subsection{Sampling Noise Control and Self-Ensemble}
\begin{figure}
    \centering
    \includegraphics[width=\linewidth]{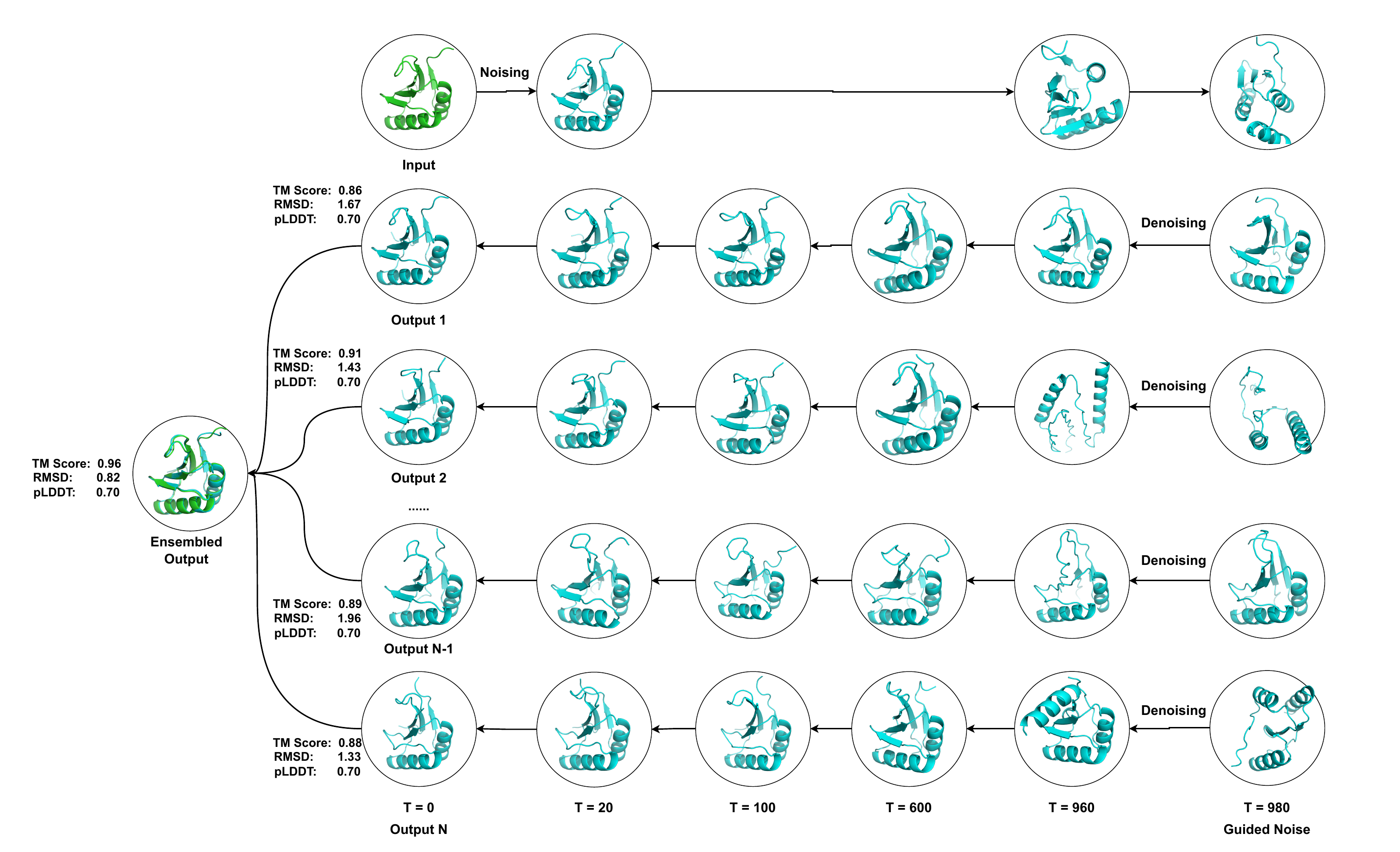}
    \caption{Illustration of the guided sampling and self-ensemble process in LaGDif. The process begins with an input protein structure, to which guided noise is applied (T=0). The model then undergoes a series of denoising steps, generating multiple output candidates. 
    These candidates are ensembled to produce the final output. 
    It shows the progression of the protein structure through various timesteps (T), with corresponding quality metrics (TM Score, RMSD, and pLDDT) for each output. 
    The ensembled output demonstrates improved structural quality (TM Score: 0.96, RMSD: 0.82) compared to individual outputs, highlighting the effectiveness of the self-ensemble method in enhancing prediction accuracy and stability.}
    \label{fig:self_ensemble}
\end{figure}

Protein inverse folding is a complex process influenced by numerous physical and chemical factors, hence denoising networks must learn to transform random noise into high-quality AA sequences while preserving spatial structure, often with limited protein data available for training. 
To address this challenge, we introduce two key innovations in our LaGDif model, namely guided sampling with controlled noise and a self-ensemble method.
we define the guided sampling process as:
\begin{equation}
\hat{x} = \alpha x + (1 - \alpha) \epsilon
\end{equation}
where $\hat{x}$ is the initial noisy node features in Eq.~\ref{eq:1}, $\alpha \in [0, 1]$ is a controllable parameter, and $\epsilon \sim \mathcal{N}(0, I)$ is Gaussian noise.
This guided sampling approach ensures diversity in the generated sequences while maintaining essential structural integrity. 
By starting from a point closer to the target distribution, we facilitate a more efficient and stable denoising process.
To further enhance the robustness and accuracy of our generated protein structures, we propose a self-ensemble method during the sampling process. 
At each denoising step $t$, we generate $K$ candidate graphs and average their node features:
\begin{equation}
x_t = \frac{1}{K} \sum_{k=1}^K f_\theta(x_{t+1}^k, E, t)
\end{equation}
where $f_\theta$ is our EGNN-based denoising network with parameters $\theta$, and $x_{t+1}^k$ is the $k$-th sampled graph node feature at step $t+1$. 
This self-ensemble method mitigates individual biases and errors that may arise during the denoising process, resulting in more reliable and consistent protein graphs.
As illustrated in Fig.~\ref{fig:self_ensemble}, the integration of guided sampling and self-ensembling allows our method to effectively balance the need for sequence diversity with structural accuracy.

\section{Experiments}
\subsection{Dataset}
\begin{figure}[htbp]
    \centering
    \includegraphics[width=\linewidth]{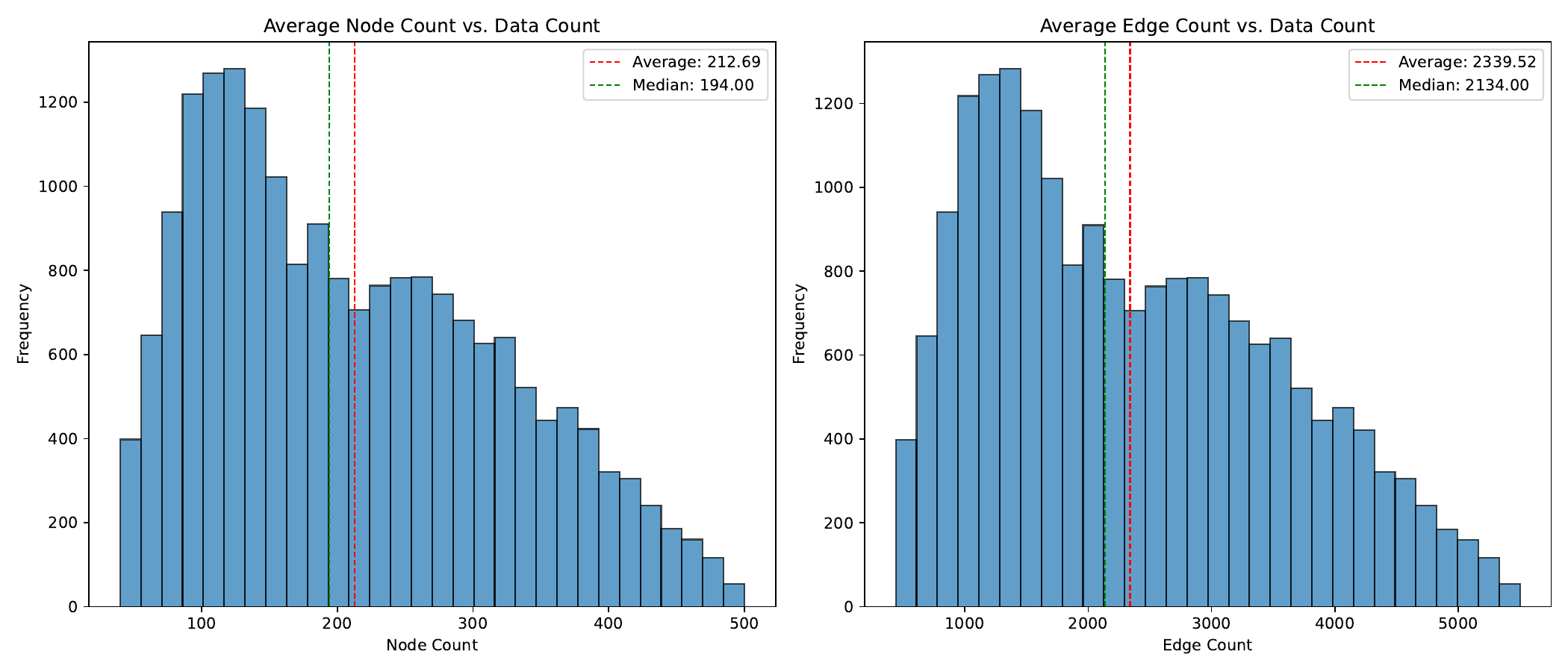}
    \caption{Distribution of protein structures in the CATH dataset used for evaluating LaGDif. 
    \textbf{Left}: Histogram of node counts (amino acids) per protein, with an average of 212.69 and median of 194.00. 
    \textbf{Right}: Histogram of edge counts (spatial connections) per protein, with an average of 2339.52 and median of 2134.00. 
    These distributions highlight the structural diversity of the dataset, ranging from short proteins to more complex structures, which allows for a comprehensive evaluation of our latent graph diffusion model across various protein sizes and complexities.}
    \label{fig:data}
\end{figure}

To evaluate the performance of LaGDif in protein inverse folding, we conducted experiments using the CATH dataset version 4.2.0 \footnote{\url{https://www.cathdb.info}}. 
%
%
%
We followed the previous data split \cite{yi2024graph}, dividing the dataset into 18,024 training samples, 608 validation samples, and 1,120 test samples. 
To assess our model's ability to handle proteins of varying complexity, we further categorized the test set into three subsets. 
The first subset consists of short proteins, comprising those with chain lengths less than 100 amino acids. 
The second subset includes single-chain proteins, consisting exclusively of proteins with a single polypeptide chain. 
The third subset encompasses all proteins in the test set, including multi-chain and longer protein structures. 
%
%
%
A distribution of the dataset statisitic is depicted in Fig.~\ref{fig:data}.

\subsection{Implementation Details}
For the denoising network in LaGDif, we utilize a 4-layer EGNN. 
We set the default number of diffusion timesteps $T$ to 1,000, striking a balance between fine-grained denoising and computational efficiency.
The LaGDif was trained for 20 epochs.
%
%
During the sampling phase, we introduced guided noise by adding perturbations to protein features at 980 timesteps, allowing for controlled exploration of the sequence space. 
To further refine our predictions, we applied a self-ensemble approach, aggregating results from five independent predictions.
Our implementation leverages PyTorch (version 1.12.1) and PyTorch-Geometric (version 2.3.1), executed on an NVIDIA\textsuperscript{\textregistered} 4090 GPU.
We utilize perplexity to quantify the model's uncertainty in predicting AA sequences, with lower values indicating more confident and accurate predictions. 
The recovery rate directly measures the model's ability to reconstruct the correct AA sequence from a given three-dimensional structure.
To evaluate structural similarity, we employ the TM-score (Template Modeling score), which assesses the topological similarity between the predicted and target protein structures on a scale from 0 to 1.
The averaged pLDDT (predicted Local Distance Difference Test) provides an estimate of local structural accuracy.
Lastly, we calculate the averaged RMSD (Root Mean Square Deviation) to measure the average distance between corresponding atoms in the predicted and target structures.%

\begin{table}[ht]
\centering
\caption{Performance comparison of LaGDif with state-of-the-art protein inverse folding models on the CATH dataset. 
Results are reported for three protein categories: short (chain length $\le$ 100), single-chain, and all proteins. 
Lower perplexity ($\downarrow$) indicates better predictive confidence, while higher recovery rate ($\uparrow$) represents more accurate sequence reconstruction. 
LaGDif demonstrates superior performance across all metrics and protein categories, showcasing its effectiveness in capturing the relationship between protein structures and sequences. The CATH version used for each model is indicated, ensuring fair comparison.}
\label{Tab:recovery}
\resizebox{\linewidth}{!}{
\begin{tabular}{lccccccccc}
\toprule
Models & \multicolumn{3}{c}{Perplexity ($\downarrow$)} & \multicolumn{3}{c}{Recovery Rate \% ($\uparrow$)} & \multicolumn{2}{c}{CATH version} \\ 
\cmidrule(lr){2-4} \cmidrule(lr){5-7} \cmidrule(lr){8-9}
 & Short & Single-chain & All & Short & Single-chain & All & 4.2 & 4.3 \\
\midrule
StructGNN \cite{ingraham2019generative} & 8.29 & 8.74 & 6.40 & 29.44 & 28.26 & 35.91 & \checkmark & \\
GraphTrans \cite{ingraham2019generative} & 8.39 & 8.83 & 6.63 & 28.14 & 28.46 & 35.82 & \checkmark & \\
GCA \cite{tan2022generative} & 7.09 & 7.49 & 6.05 & 32.62 & 31.10 & 37.64 & \checkmark & \\
GVP \cite{jing2020learning} & 7.23 & 7.84 & 5.36 & 30.60 & 28.95 & 39.47 & \checkmark & \\
GVP-large \cite{hsu2022learning} & 7.68 & 6.12 & 6.17 & 32.60 & 39.40 & 39.20 &  & \checkmark \\
AlphaDesign \cite{gao2022alphadesign} & 7.32 & 7.63 & 6.30 & 34.16 & 32.66 & 41.31 & \checkmark & \\
ESM-IF1 \cite{hsu2022learning} & 8.18 & 6.33 & 6.44 & 31.30 & 38.50 & 38.30 &  & \checkmark\\
ProteinMPNN \cite{dauparas2022robust} & 6.21 & 6.68 & 4.57 & 36.35 & 34.43 & 49.87 & \checkmark & \\
PIFold \cite{gao2022pifold} & 6.04 & 6.31 & 4.55 & 39.84 & 38.53 & 51.66 & \checkmark & \\
Grade-IF \cite{yi2024graph} & 5.49 & 6.21 & 4.35 & 45.27 & 42.77 & 52.21 & \checkmark & \\
\midrule
LaGDif (Ours) & \underline{\textbf{2.70}} & \underline{\textbf{2.56}} & \underline{\textbf{2.46}} & \underline{\textbf{86.97}} & \underline{\textbf{88.32}} & \underline{\textbf{88.73}} & \checkmark & \\
\bottomrule
\end{tabular}
}
\end{table}

\begin{table}[ht]
\centering
\caption{Structural quality assessment of protein sequences generated by various inverse folding models. The comparison uses three key metrics: TM-score (higher is better, max 1.0), average pLDDT (higher is better, max 1.0), and average RMSD (lower is better, in \AA). 
Results are presented as mean $\pm$ standard deviation, demonstrating LaGDif's competitive performance in maintaining structural integrity while performing inverse folding.}
\label{Tab:numerical_comparison}
\begin{tabular}{lccc}
\toprule
Methods  & TM score & avg pLDDT & avg RMSD \\
\midrule
GVP  & 0.24 {\tiny $\pm$ 0.01} & 0.74 {\tiny $\pm$ 0.04} & 4.66 {\tiny $\pm$ 0.01} \\
ESM-IF1  & 0.81  {\tiny $\pm$ 0.01} &  0.72 {\tiny $\pm$ 0.04} &  1.97 {\tiny $\pm$ 0.03} \\
PIFold  & 0.68 {\tiny $\pm$ 0.01} & 0.63 {\tiny $\pm$ 0.01} & 2.54 {\tiny $\pm$ 0.05} \\
ProteinMPNN  & 0.84 {\tiny $\pm$ 0.03} & 0.79 {\tiny $\pm$ 0.03} & 1.76 {\tiny $\pm$ 0.03} \\
Grade-IF & 0.78 {\tiny $\pm$ 0.01} & 0.70 {\tiny $\pm$ 0.03} & 2.17 {\tiny $\pm$ 0.03} \\
\hline
LaGDif (Ours) & 0.82 {\tiny $\pm$ 0.01}  & 0.70 {\tiny $\pm$ 0.01}  &  1.96 {\tiny $\pm$ 0.03} \\
\bottomrule
\end{tabular}
\end{table}

\subsection{Results}
Our evaluation of LaGDif focused on two aspects, namely the sequence recovery performance and structural quality of the generated proteins.
As demonstrated in Table \ref{Tab:recovery}, LaGDif outperformed existing state-of-the-art methods in the inverse folding task across all protein categories. 
Our model achieved the following improvements in recovery rates: $41.7\%$ for short-chain proteins ($86.97\%$ vs. $45.27\%$ for Grade-IF), $45.55\%$ for single-chain proteins ($88.32\%$ vs. $42.77\%$), and $36.52\%$ for all tested proteins ($88.73\%$ vs. $52.21\%$). 
These improvement in recovery rates can be attributed to our novel approach combining latent space diffusion, guided initial sampling noise, and self-ensemble scheme. 
Moreover, LaGDif demonstrated superior predictive confidence, as evidenced by significantly lower perplexity scores across all protein categories. The perplexity values for LaGDif ($2.70$, $2.56$, and $2.46$ for short, single-chain, and all proteins, respectively) are lower than those of the next best performer, Grade-IF ($5.49$, $6.21$, and $4.35$), indicating a more certain and accurate prediction process.
To evaluate the reliability of sequences generated by LaGDif, we assessed the structural similarity between the predicted sequences and the original native structures after folding. 
We used ESMFold \cite{lin2023evolutionary} to generate 3D structures for both predicted and original sequences, then compared their structural properties. 
As shown in Table \ref{Tab:numerical_comparison}, LaGDif demonstrated competitive performance in maintaining structural integrity. 
Our method achieved an average TM-score of $0.82 \pm 0.01$, slightly lower than ProteinMPNN ($0.84 \pm 0.03$) but higher than other methods. 
This score, well above the $0.5$ threshold, indicates a high degree of structural similarity between generated and native proteins. 
In terms of average pLDDT, our method attained a score of $0.70 \pm 0.01$, on par with Grade-IF and slightly lower than ProteinMPNN ($0.79 \pm 0.03$), suggesting good local structural accuracy in our generated proteins. 
Notably, LaGDif achieved an average RMSD of $1.96 \pm 0.03$ $\text{\AA}$, second only to ProteinMPNN ($1.76 \pm 0.03$ $\text{\AA}$). 
For high-resolution crystal structures, an RMSD less than $2$ $\text{\AA}$ is considered very close, indicating that our method produces structures highly similar to the native ones.
These results demonstrate that LaGDif not only excels in sequence recovery but also maintains high structural fidelity in the generated proteins. 

\subsection{Ablation Study}

\begin{figure}
    \centering
    \includegraphics[width=\linewidth]{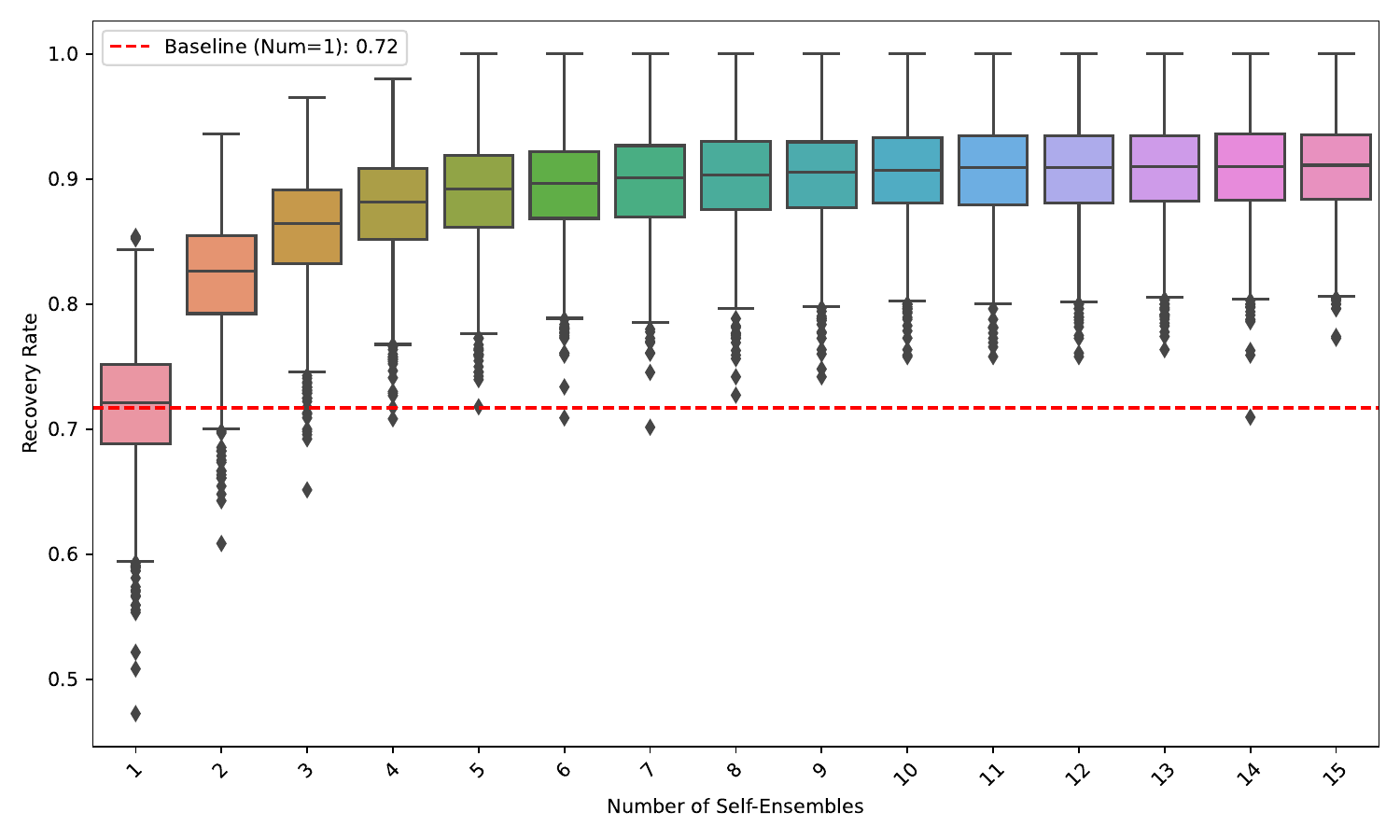}
    \caption{Impact of $K$ in self-ensemble on LaGDif's recovery rate performance. 
    The graph shows the recovery rate improvement as the number of self-ensemble iterations increases from 1 to 15. 
    The baseline recovery rate (0.72) is achieved with a single iteration ($K=1$) \textit{i}.\textit{e}., no self-ensemble. 
    The plot demonstrates a significant improvement in recovery rate up to 5-7 iterations, after which the gains begin to plateau, illustrating the effectiveness and diminishing returns of the self-ensemble in enhancing inverse folding performance.}
    \label{fig:self_ensemble_performance}
\end{figure}

To assess the impact of key components in LaGDif, we conducted a ablation study. Our experiments focused on two aspects, namely the number of $K$ in self-ensemble and the guided initial noise.
As illustrated in Table \ref{Tab:ablation}, the self-ensemble approach enhances the model's performance by mitigating prediction errors. 
We observed a consistent improvement across all metrics as the number of ensemble samples ($K$) increased. 
Notably, the recovery rate improved from $0.7216$ for the baseline model (sampling from noisy graph) to $0.8904$ with $K=5$ in self-ensemble, demonstrating a substantial $16.88\%$ increase.
To further investigate the impact of $K$ in self-ensemble iterations on recovery rate, we extended our analysis from $1$ to $15$, as shown in Fig. \ref{fig:self_ensemble_performance}. 
The results reveal an observable improvement in recovery rate, particularly between $5$-$7$ iterations, after which the gains begin to plateau. 
This trend underscores the effectiveness of the self-ensemble technique in enhancing inverse folding performance, with the median recovery rate steadily increasing and stabilizing after multiple iterations. 
Importantly, all self-ensemble configurations consistently outperformed the baseline recovery rate of $0.72$ (at $K=1$), validating the robustness of our approach.

\begin{table}[h!]
    \centering
    \caption{Ablation study demonstrating the impact of $K$ in self-ensemble, and guided noise on LaGDif's performance.
    Results show the progressive improvement in all metrics as the number of $K$ in self-ensemble layers increases. 
    The baseline models, sampling from noisy graph and pure noise, illustrate the important role of guided noise in enhancing model performance. 
    pLDDT, TM-score, and RMSD are presented as mean $\pm$ standard error, providing insight into the consistency and reliability of the results.}
    \resizebox{\linewidth}{!}{
    \begin{tabular}{lcccc}
        \toprule
        Methods & Recovery Rate & pLDDT & TM Score & RMSD \\
        \midrule
        Self-Ensemble ($K=5$) & 0.8904 & 69.82 {\tiny $\pm$ 0.39} & 0.82 {\tiny $\pm$ 0.005} & 1.96 {\tiny $\pm$ 0.03} \\
        Self-Ensemble ($K=4$) & 0.8807 & 68.98 {\tiny $\pm$ 0.40} & 0.81 {\tiny $\pm$ 0.005} & 2.04 {\tiny $\pm$ 0.03} \\
        Self-Ensemble ($K=3$) & 0.8632 & 67.93 {\tiny $\pm$ 0.39} & 0.80 {\tiny $\pm$ 0.005} & 2.15 {\tiny $\pm$ 0.03} \\
        Self-Ensemble ($K=2$) & 0.8257 & 64.60 {\tiny $\pm$ 0.42} & 0.75 {\tiny $\pm$ 0.006} & 2.40 {\tiny $\pm$ 0.04} \\
        Baseline Model (Sampling from Noisy Graph) & 0.7216 & 56.59 {\tiny $\pm$ 0.43} & 0.66 {\tiny $\pm$ 0.006} & 2.97 {\tiny $\pm$ 0.04} \\
        Baseline Model (Sampling from Pure Noise) & 0.3170 & 39.59 {\tiny $\pm$ 0.27} & 0.28 {\tiny $\pm$ 0.001} & 4.88 {\tiny $\pm$ 0.03} \\
        \bottomrule
    \end{tabular}
    \label{Tab:ablation}
    }
\end{table}

\begin{figure}[htbp]
    \centering
    \includegraphics[width=\linewidth]{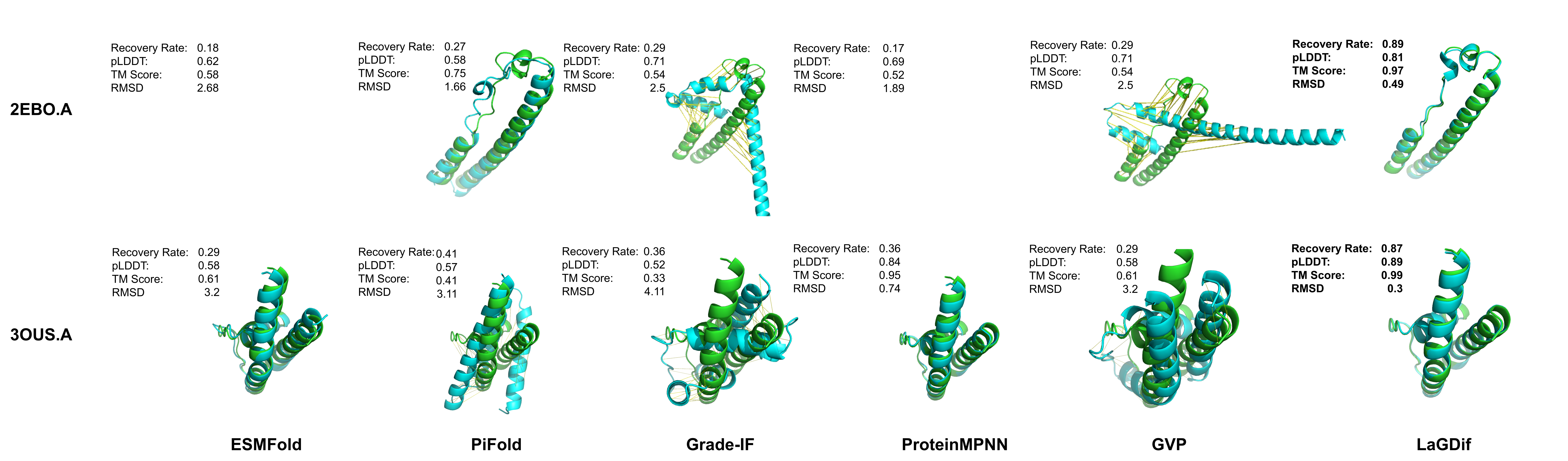}
    \caption{
    %
    %
    Visualization of the generated structures and performance metrics on the two case studies. 
    The top row shows results for the 2EBO protein (Ebola virus GP2 core structure), while the bottom row presents results for the 3OUS protein (mutant MthK channel pore). 
    LaGDif consistently outperforms other models, achieving the highest recovery rates (0.89 for 2EBO.A and 0.97 for 3OUS.A) and demonstrating superior structural fidelity as indicated by improved TM Scores and RMSD values.
    }
    \label{fig:case_study}
\end{figure}

\subsection{Case Study}
To provide a more detailed assessment of LaGDif's performance, we conducted in-depth case studies on two specific proteins, namely 2EBO and 3OUS, as shown in Fig.~\ref{fig:case_study}. 
The 2EBO protein represents the core structure of the Ebola virus GP2, which is important for viral and host membrane fusion. 
In contrast, 3OUS is a mutant of the MthK channel pore from Methanothermobacter thermautotrophicus. 
This potassium ion channel, with its complex structure resolved by high-resolution X-ray diffraction at $1.75$ $\text{\AA}$, is important for studying ion selectivity and membrane transport mechanisms.
We compared LaGDif's performance on the A chain of these proteins against several state-of-the-art protein inverse folding models, including ESM-IF, PiFold, GradeIF, ProteinMPNN, and GVP. As illustrated in Fig.~\ref{fig:case_study}, LaGDif consistently achieved the highest metrics. 
Moreover, when the predicted sequences were folded using ESMfold \cite{lin2023evolutionary}, they demonstrated a high degree of structural consistency with the original sequences.
For the 2EBO protein, LaGDif achieved remarkable results with a recovery rate of $89\%$, an RMSD of $0.49$ $\text{\AA}$, and a TM score of $0.87$. 
The more complex 3OUS protein showcased LaGDif's exceptional performance, achieving a recovery rate of $87\%$, an RMSD of $0.3$ $\text{\AA}$, and a TM score of $0.99$. 
These results consistently outperformed other leading models across all metrics. 
The LaGDif's ability to excel with proteins of varying complexities, from the relatively simpler 2EBO to the more intricate 3OUS, underscores its robustness and versatility.

\subsection{Model Complexity Analysis}

Table \ref{tab:model_performance} presents a comparison of model parameters, inference time, and memory usage across different approaches.
LaGDif demonstrates a balanced profile in terms of model complexity.
With approximately $20.351$ million parameters, it sits in the middle range of model sizes, smaller than ESM-IF ($141.662$M) but larger than ProteinMPNN ($1.660$M) and Grade-IF ($3.818$M).
This moderate parameter count allows LaGDif to capture complex protein structures without excessive computational overhead.
In terms of inference time, LaGDif requires $2.95$ seconds, which is longer than ProteinMPNN ($0.45$s) and Grade-IF ($0.37$s), but considerably faster than GVP ($17.50$s). 
This performance indicates that while LaGDif may not be the fastest model, it offers a reasonable trade-off between speed and accuracy, especially considering its superior performance in sequence recovery and structural fidelity as demonstrated in our earlier experiments.
Memory usage is another important factor in model deployment. 
LaGDif utilizes $280.97$ MB, which is higher than ProteinMPNN ($12.79$ MB) and Grade-IF ($29.14$ MB) but substantially lower than GVP ($4614.60$ MB) and ESM-IF ($2374.59$ MB). 

\begin{table}[ht]
    \centering
    \caption{Evaluation on the model complexity in terms of the number of parameters (in millions), inference time (in seconds), and memory usage (in megabytes). }
    \resizebox{0.8\linewidth}{!}{
    \begin{tabular}{@{}lcccc@{}}
        \toprule
        Models &
        \# Params. (M) & Time (s) &  Memory (MB)  \\

        \midrule
        GVP & 1,010,920 & 17.50 & 4614.60\\

        ProteinMPNN & 1.660 & 0.45 & 12.79\\

        ESM-IF & 141.662 & 1.39 & 2374.59 \\

        Grade-IF & 3.818 & 0.37 & 29.14\\
        \hline
        LaGDif & 20.351 & 2.95 & 280.97 \\
        \bottomrule
    \end{tabular}
    }
    \label{tab:model_performance}
\end{table}

\section{Conclusion}

In this work, we introduced LaGDif, a novel latent graph diffusion model for protein inverse folding that bridges discrete and continuous spaces. 
By leveraging the power of diffusion models in continuous domains, LaGDif addresses the limitations of existing discrete diffusion approaches and traditional language model-based methods in protein sequence generation.
Our comprehensive evaluation on the CATH dataset demonstrates that LaGDif outperforms state-of-the-art inverse folding methods across various protein categories. 
The advancements presented in LaGDif have implications for the field of protein design and engineering. By enabling more accurate and diverse protein sequence generation, our approach has the potential to accelerate the development of novel proteins for therapeutic and industrial applications. 
Future work could explore the application of LaGDif to more challenging protein design tasks, such as de novo protein design or protein-protein interaction prediction.

\bibliographystyle{IEEEtran}
\bibliography{reference}
\end{document}